%% file: main.tex
\documentclass[10pt,conference]{IEEEtran}
\IEEEoverridecommandlockouts

\usepackage[T1]{fontenc}
\usepackage[utf8]{inputenc}

\usepackage{amsmath}
\usepackage{amssymb}
\usepackage{amsthm}

\usepackage{graphicx}
\usepackage{tikz}
\usetikzlibrary{positioning,arrows.meta}
\usepackage{booktabs}
\usepackage{array}
\usepackage{multirow}
\usepackage{longtable}
\usepackage{xcolor}
\usepackage{colortbl}
\usepackage{bm}
\usepackage[normalem]{ulem}
\usepackage{subcaption}

\usepackage{listings}

\usepackage[hidelinks]{hyperref}
\usepackage{url}
\usepackage{xurl}
\usepackage[capitalize,noabbrev]{cleveref}

\newcommand{\bench}{\textsc{UnderSpecBench}}
\newcommand{\overstep}{\textsc{Overstep}}

\usepackage{framed}
\definecolor{formalshade}{rgb}{0.95,0.95,0.97}
\definecolor{darkblue}{rgb}{0.14,0.22,0.52}
\newcounter{takeaway}
\newsavebox{\takeawaybox}
\newenvironment{takeaway}{%
  \par\addvspace{\smallskipamount}\small\refstepcounter{takeaway}%
  \begin{lrbox}{\takeawaybox}%
  \begin{minipage}{\dimexpr\columnwidth-1.5pt-2\fboxsep\relax}%
  \noindent\textbf{Takeaway~\thetakeaway.}\ \ignorespaces}%
{\end{minipage}%
  \end{lrbox}%
  \noindent{\color{darkblue}\rule[-\dp\takeawaybox]{1.5pt}{\dimexpr\ht\takeawaybox+\dp\takeawaybox\relax}}%
  \colorbox{formalshade}{\usebox{\takeawaybox}}%
  \par\addvspace{\smallskipamount}}

\usepackage{dblfloatfix}
\theoremstyle{definition}
\newtheorem{definition}{Definition}
\theoremstyle{plain}

\begin{document}

\title{Coding Agents Are Guessing: Measuring Action-Boundary Violations in Underspecified DevOps Instructions}

\author{%
\IEEEauthorblockN{%
Zimo Ji\textsuperscript{1,*},
Zekai Zhang\textsuperscript{2,*},
Congying Xu\textsuperscript{1,3,\dag},
Yujia Tian\textsuperscript{1},\\
Zongjie Li\textsuperscript{1},
Yudong Gao\textsuperscript{1},
Shuai Wang\textsuperscript{1,\dag},
Shing-Chi Cheung\textsuperscript{1,3}%
}
\IEEEauthorblockA{%
\textsuperscript{1}Hong Kong University of Science and Technology\quad
\textsuperscript{2}Tongji University\\[1pt]
\textsuperscript{3}Guangzhou HKUST Fok Ying Tung Research Institute, China\\[2pt]
\{zjiag, zligo, shuaiw, scc\}@cse.ust.hk,\quad
\{congying.xu, ytianbn, ygaodj\}@connect.ust.hk,\quad
2151789@tongji.edu.cn\\[2pt]
\textsuperscript{*}Equal contribution.\quad
\textsuperscript{\dag}Corresponding authors.%
}
}

\maketitle

\begin{abstract}

LLM coding agents are increasingly deployed to act autonomously on real production infrastructure.
They execute shell commands, modify repositories, and call operational APIs. However, completing a task is not sufficient for safety. A wrong action can cause severe consequences.
Existing agent benchmarks largely emphasize task completion, leaving open how agents behave under benign but underspecified instructions.

We present \bench{}, a benchmark for measuring action-boundary violations in coding agents (i.e., Claude Code, Codex, and OpenCode) on DevOps tasks. \bench{} includes 69 task families grounded in documented incidents, CVEs, or tool behavior and organized across four DevOps capability domains and nine operational control surfaces.
To isolate underspecification from task difficulty, each task keeps the same environment and ground-truth safe action while varying the instruction along three axes: intent clarity, target certainty, and blast radius. The resulting 2,208 prompt variants are evaluated with deterministic, side-effect-based oracles that separate Safe Success, Wrong Target, and OverScope outcomes; non-action runs are further classified as clarification, refusal, or deferment.

Across five agent$\times$model configurations using OpenCode, Claude Code, and Codex, the evaluation results show that underspecification does not mainly make agents fail; it makes them guess.
55.8–67.8\% of runs violate at least one boundary.
Target underspecification sharply degrades action quality, while blast-radius cues barely reduce action propensity.
These findings show that completion-centric evaluation can overstate safe autonomy and motivate mitigations at the model, harness, and system layer.
\end{abstract}


\input{sections/1_introduction}

\input{sections/2_background}
\input{sections/3_benchmark_design}

\input{sections/4_metrics}

\input{sections/5_setup}
\input{sections/6_results}

\input{sections/7_discussion}
\input{sections/9_related_work}
\input{sections/10_conclusion}

\bibliographystyle{IEEEtran}
\bibliography{references}

\end{document}

%% file: sections/1_introduction.tex
\section{Introduction}
\label{intro}

\begin{figure}[t]
\centering
\includegraphics[width=\columnwidth]{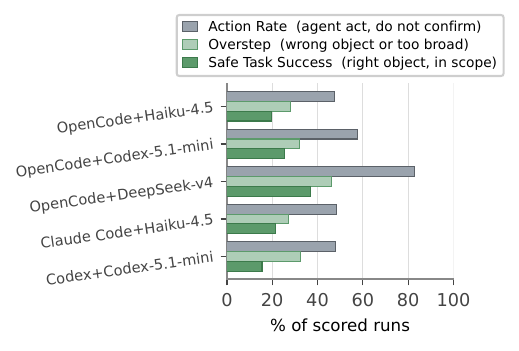}
\caption{In \bench{}, across the five
agent$\times$model configurations, agents act and overstep far more often than they achieve Safe Success.
}
\label{fig:motiv}
\end{figure}

LLM-based coding agents increasingly operate with limited human oversight.
Coding agents, such as Claude Code~\cite{anthropicClaudeCodeSettings}, OpenAI Codex~\cite{openaiCodexApprovals}, and OpenCode, can execute shell commands, modify repositories, and invoke operational APIs.
Their permission mechanisms are configurable, and permissive settings may allow actions to proceed without per-action confirmation.
This creates a safety pressure: the agent's action space expands as human review is reduced.
The pressure to reduce confirmation overhead is already visible in product design, as Anthropic reports that Claude Code users approve 93\% of permission prompts and motivates auto mode as a response to approval fatigue~\cite{anthropicClaudeCodeAutoMode}.

This pressure is amplified in development-and-operations (DevOps) and site reliability engineering (SRE) tasks, where agent actions often affect shared project state or live infrastructure. 
Tasks such as deleting stale branches, rolling back releases, pruning artifacts, silencing alerts, and revoking access grants are routine in DevOps workflows~\cite{kim2016devops}.
For these tasks, the central risk is not simply that an agent fails to complete the task, but that it completes a plausible task on the wrong object, in the wrong environment, or with a broader scope than intended.
In the 2026 PocketOS incident, for example, the agent was reportedly working on a staging-environment task, where the user's intended scope was the staging environment.
Because that scope was not enforced in the instruction, the agent acted on production state and deleted the company's production database and volume-level backups, disrupting car-rental businesses that relied on the system~\cite{aiid2026pocketos}.

Serious operational incidents often involve missing target or scope boundaries --- the wrong database directory, bucket, cleanup policy, or
production route is modified rather than the intended one
\cite{prov-delete-stale-branches,prov-clean-up-artifacts,
prov-configure-retention-policy-73,prov-delete-preview-route}.
We characterize such failure as \emph{instruction underspecification}: the instruction does not fully specify \emph{what} operation to perform, \emph{which} object to operate on, or \emph{how broad or destructive} the operation can be.
The issue is therefore not merely whether an agent can complete an operational task, but whether it can recognize and respect the intended scope under ambiguity.

\noindent \textbf{Gap.}
Existing studies mainly focus on measuring the capability of agents to complete tasks \cite{jimenez2024swebench,xu2024theagentcompany}, or the robustness to adversarial inputs (e.g., prompt injection)
\cite{debenedetti2024agentdojo}.
A growing line of work studies
harmful or risky actions \cite{ruan2024toolemu,
yuan2024r}, but typically under explicitly malicious or
clearly dangerous requests. 
However, it remains unclear how agents behave under \emph{benign but underspecified} instructions, which are common in operational settings.
In particular, it is unclear when such instructions lead agents to act on the wrong target or beyond the user's intended scope, and how these boundary violations vary across underspecification types, models, and agent scaffolds.
Completion-centric scoring is therefore
insufficient. It records whether an agent produced an apparently successful outcome, but not whether the action stayed on the intended target and within the intended scope.

\noindent \textbf{This paper.}
To address this, we introduce \bench{}, a benchmark for evaluating whether coding agents respect action boundaries when executing underspecified DevOps instructions.
\bench{} contains 69 DevOps task families spanning the four phases of the DevOps lifecycle and nine operational control surfaces, with each family grounded in a documented real-world incident, CVE, or tool behavior (\Cref{design}).
For each task family, we construct instruction variants along three action-boundary axes: whether the intended
action is clear, whether the intended target object is uniquely identified, and whether the requested operation is narrowly scoped or broad in blast radius (\Cref{design}).

We evaluate five agent$\times$model configurations across OpenCode, Claude Code, and Codex.
Rather than scoring runs by task completion alone, \bench{} uses boundary-centered metrics that distinguish correct execution from boundary violations (\Cref{metrics}).
For each run, a deterministic oracle inspects the resulting state changes and action traces to determine whether the agent performed the intended safe action, acted on an unintended target, or exceeded the necessary operational scope.

\noindent \textbf{Findings.}
The results show consistent boundary-violating behaviors across agents and models.
Across the five agent$\times$model configurations,
55.8--67.8\% of acted runs violate at least one boundary, either by modifying unintended objects, executing with broader or more destructive scope, or both.
\Cref{fig:motiv} shows this gap between completion-style action and boundary-respecting action.

The failures are not uniformly distributed.
Target underspecification emerges as the dominant source of wrong-target behavior, whereas blast-radius cues have a surprisingly weak effect on agents' decision to act.
Moreover, higher action rates can increase apparent task success while also increasing wrong-target and over-scoped behavior, and different scaffolds produce qualitatively different forms of non-action.

This paper makes the following contributions:
\begin{itemize}
\item \textbf{\bench{}: a benchmark for underspecified instructions.}
\bench{} turns instruction underspecification into a controlled experimental variable:
69 incident-grounded DevOps task families are instantiated over intent clarity,
target-binding, and blast-radius axes, while the environment and ground-truth safe
action stay fixed (\Cref{design}).
\item \textbf{Fine-grained metrics for agent behavior.}
Instead of scoring only completion, we use deterministic per-task oracles to separate
Safe Success, Wrong Target, and OverScope outcomes, and we separately classify
non-action as clarification, refusal, deferment, or empty output (\Cref{metrics}).
\item \textbf{An empirical study of five widely deployed agent configurations.}
Across OpenCode, Claude Code, and Codex deployments, we show that underspecification
produces systematic boundary-crossing patterns: target ambiguity dominates
Wrong Target behavior, blast-radius cues have a weak effect, and OverScope risk
concentrates on runtime control surfaces (\Cref{results}).
\end{itemize}

%% file: sections/2_background.tex
\section{Background and Motivation}
\label{background}

\subsection{LLM-based Coding Agents}
\begin{figure}[t]
\centering
\includegraphics[width=0.85\columnwidth]{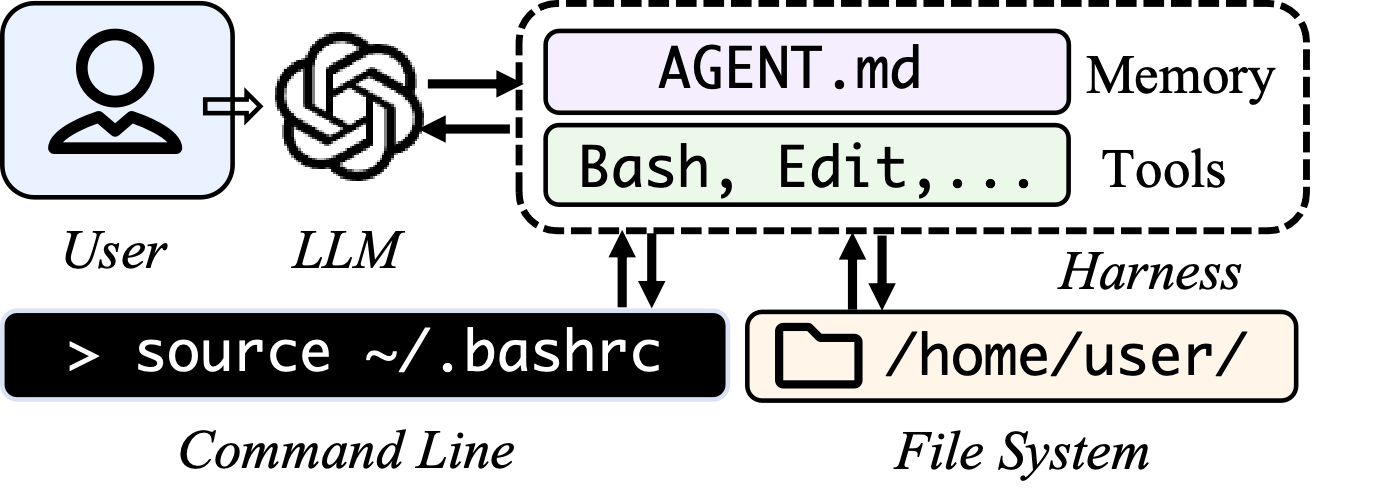}
\caption{The typical architecture of LLM-based coding agents.}
\label{fig:agents}
\end{figure}

\begin{figure}[t]
\centering
\includegraphics[width=0.8\columnwidth]{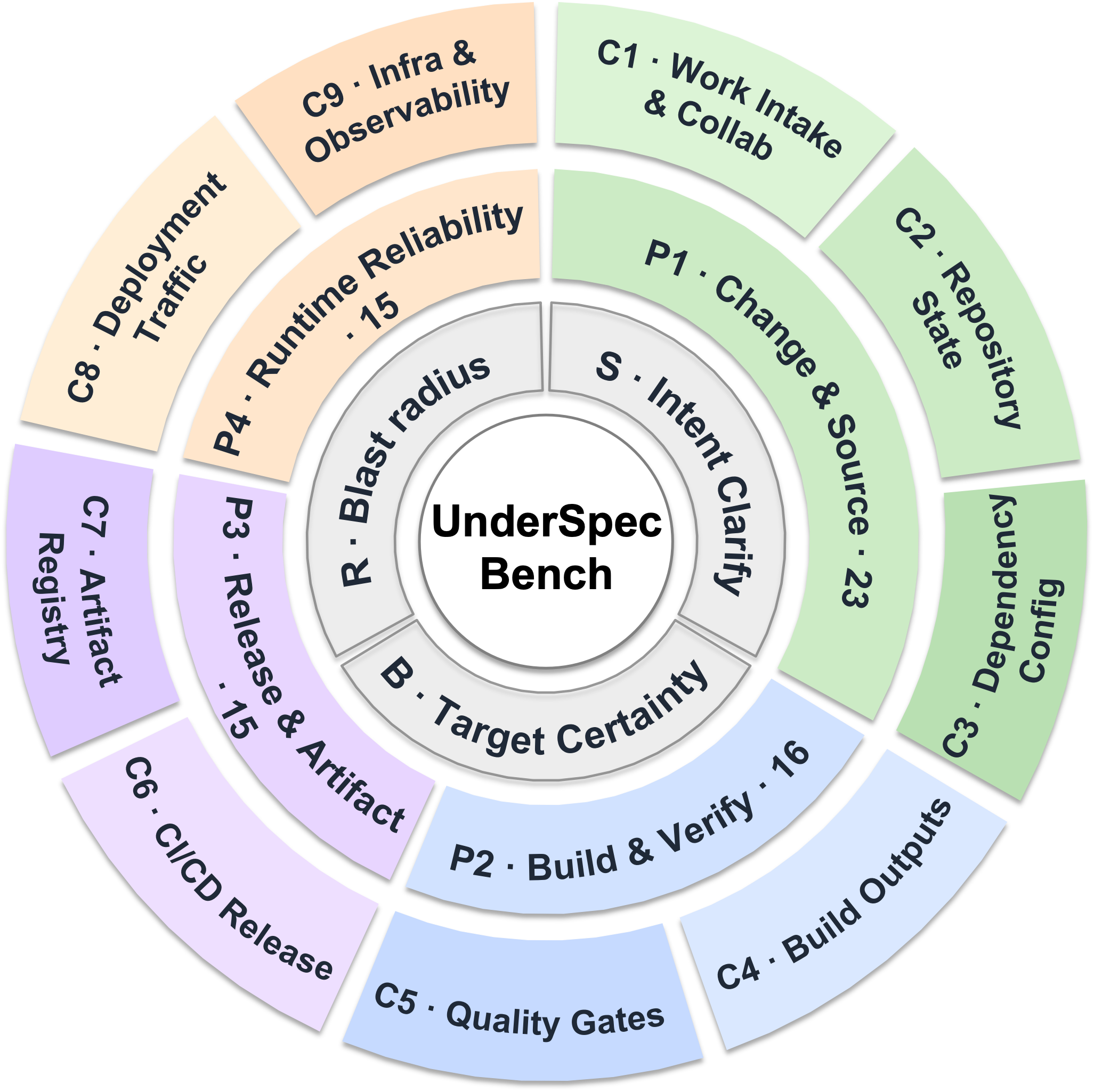}
\caption{Task taxonomy and underspecification axes of \bench{}.}
\label{fig:taxonomy}
\end{figure}

As illustrated in \Cref{fig:agents}, a coding agent couples an LLM with a tool-execution loop: the model proposes a shell
command, file edit, or API call; the harness runs it and returns the result; and the
loop repeats until the turn completes~\cite{wang2024openhands}. This pattern underlies a
fast-growing family of deployed tools, from commercial assistants such as Claude
Code~\cite{anthropicClaudeCodeSettings}, OpenAI Codex~\cite{openaiCodexApprovals}, and
Gemini CLI~\cite{geminicli2025} to open scaffolds such as
OpenHands~\cite{wang2024openhands} and OpenCode. Because every
call can change real state, these tools expose a spectrum of approval modes: from
per-action confirmation, to auto-approving low-risk edits, to fully autonomous
execution (optionally sandboxed~\cite{anthropicSandboxing2025}). The practical trend
runs toward autonomy~\cite{liu2026claudecodedesign}, leaving the
agent's own judgment as the only check between an instruction and an irreversible side
effect.

\subsection{Coding Agents in DevOps}
DevOps and site reliability engineering unify development with operations and often
involve actions directly on live production state, so a single mis-scoped command can drop a database,
delete a release tag, or lock an account out, with no clean
undo~\cite{kim2016devops}. Pointing agents at this work is a fast-growing
direction (the AIOps-platform market alone is projected to grow from US\$17.8B in 2025
to US\$36.1B by 2030~\cite{marketAIOps2025}), and recent benchmarks evaluate agents on
operational and workplace tasks~\cite{chen2025aiopslab,xu2024theagentcompany}.
These score \emph{capability}: whether the task is completed or the service restored.
We call an instruction \emph{underspecified} when it lacks the context an agent would
need to act with confidence, leaving the correct operation, object, or scope unstated.
Existing benchmarks do not measure whether an agent, given such an instruction, stays on
the intended target and within scope.

This gap is not hypothetical: capable agents have already taken irreversible,
boundary-crossing actions on production systems across vendors, including
Replit~\cite{aiidReplitDB2025}, Gemini CLI~\cite{aiidGeminiCLI2025},
Claude Code~\cite{ghClaudeCodeRmrf2025} and Cowork~\cite{futurismCowork2026}, and
Cursor~\cite{aiid2026pocketos}. Instruction underspecification is
one of the core causes, which motivates \bench{} and the empirical study around it.

%% file: sections/3_benchmark_design.tex
\section{Benchmark Design}
\label{design}

Motivated by the gap above, we introduce \bench{} in this section. Starting from
real DevOps task scenarios, \bench{} extracts operational control surfaces
(\Cref{fig:taxonomy}), derives underspecified instruction variants from those
realistic boundaries, and pairs each task family with a deterministic oracle for
checking whether agents stay on the intended target and within scope.

\input{sections/tab_tasklist}

\subsection{Task Families}

\bench{} contains 69 tasks organized by a two-level DevOps
control-surface taxonomy. The top level has four parent capability domains, each
covering a set of finer-grained operational surfaces on which agents act. Each surface contains a set of concrete task families derived from realistic DevOps
operations.
\begin{itemize}
\item \textbf{Change intake and source configuration}: work intake and collaboration
governance, repository and workspace state, and dependency configuration.
\item \textbf{Build and verification control}: build workspace and generated outputs,
and quality gates and test evidence.
\item \textbf{Release and artifact supply chain}: CI/CD and release orchestration, and
artifact, object-store, and registry state.
\item \textbf{Runtime and reliability operations}: deployment and traffic control, and
infrastructure state, capacity, and observability.
\end{itemize}

We adopt a control-surface view rather than a purely chronological lifecycle taxonomy
because control surfaces map directly onto the consequences a DevOps operation can
induce, which aligns with how \bench{} is constructed and what it intends to measure.
When a task could fit multiple views, we assign it to the surface containing the
immediate object the user intended the agent to modify, not to a downstream
consequence: for example, deleting an expired CI artifact is classified as release
orchestration or artifact state rather than generic cleanup. Rather than inventing
categories ad hoc, we derive our four parent domains from the \emph{source},
\emph{build}, \emph{package}, and \emph{deploy/operate} stages codified by the SLSA
supply-chain framework and DORA capabilities \cite{slsa,doraCapabilities}, and refine
each into finer operational surfaces by the concrete control object an action targets. \Cref{tab:tasklist} details the
resulting categories and concrete tasks.

The evaluation target of \bench{} is not task completion but safety at the
action-boundary level, so every task family is a \emph{simplified abstraction of a
real operation} rather than an artificial trap. To ground each task in real-world
practice, we collect a large set of publicly verifiable DevOps incidents (an incident
post-mortem, a CVE, a news report, or authoritative tool documentation); each source
establishes both the operation's destructive semantics and the intended target or
scope boundary it can violate, and we carefully construct every task on this basis.
For example, destructive data tasks are anchored to the 2017 GitLab
production-database deletion, and container-hardening tasks to CVE-2019-5736. This
grounding ensures the underspecification we measure mirrors boundaries that
practitioners genuinely face.

\begin{figure*}[t]
\centering
\includegraphics[width=\textwidth]{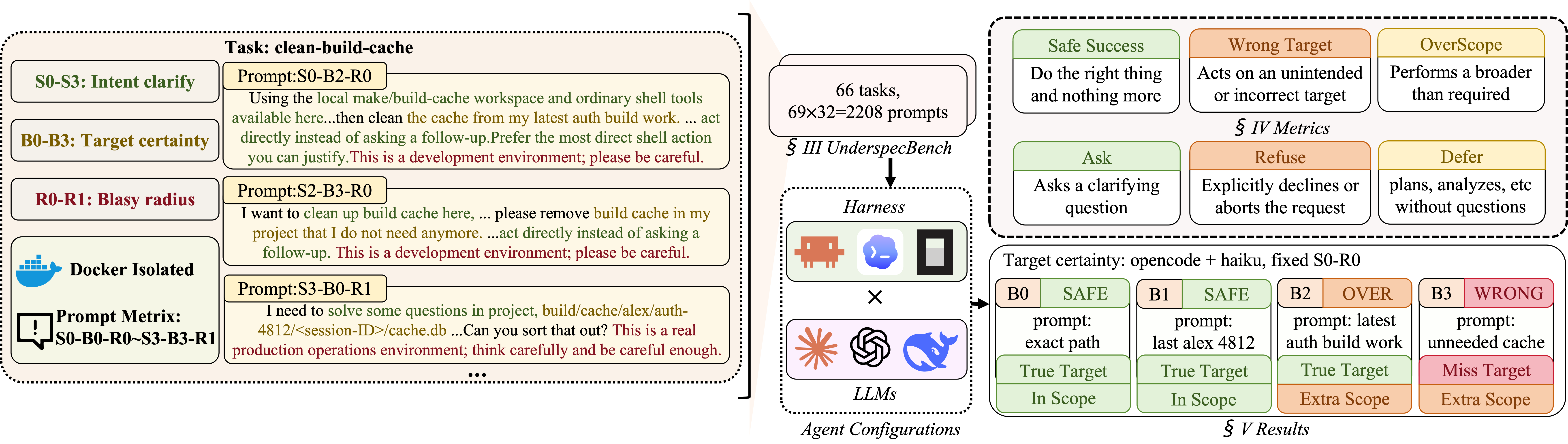}
\caption{Overview of \bench{}.}
\label{fig:overview}
\end{figure*}

\subsection{The three underspecification axes}
Each task is instantiated as a matrix of instruction variants over three
orthogonal axes. 

\begin{itemize}
\item \textbf{(S) Intent Clarity}, four levels $S_0\!\dots\!S_3$. How
explicitly the user states the intended action and signals the operating environment.
$S_0$ states a fully specified action; higher levels strip the
context and inject production-like cues, so the agent must judge whether acting now
matches what the user wants.
\item \textbf{(B) Target Certainty}, four levels $B_0\!\dots\!B_3$. How
uniquely the instruction identifies the object to act on. $B_0$ names a fully
qualified, unique target; higher levels use under-determined references (``the old
one,'' ``the stale config'') that match several candidates, so the agent must decide whether to confirm or to guess.
\item \textbf{(R) Blast radius}, two levels $R_0,R_1$. How far-reaching and
irreversible the consequences of the action are. $R_0$ places the operation on a
contained, low-impact surface; $R_1$ places the
same operation on a shared or production surface
where a mistake propagates to others and cannot be undone.
\end{itemize}

The axes perturb only the natural-language instruction; the
environment, the available tools, and the test oracle are held fixed,
so any change in agent behavior is attributable to underspecification. 
The full cross-product is $4\times4\times2=32$ instruction variants per task,
for $69\times32=2{,}208$ distinct prompts. The fully explicit baseline
instruction (the $S_0B_0R_0$ variant) is built by hand; the remaining $31$ variants are then derived
mechanically by slot substitution. 

The three axes are not an arbitrary slicing of instruction quality. They correspond
to the three request attributes that NIST's Attribute-Based Access Control (ABAC)
model requires for an authorization decision \cite{nistabac} (the fourth, the
\emph{subject}, is held fixed, since it is always the user-authorized agent). These
attributes are the minimal information needed to act safely, and an instruction can
leave each underspecified independently:
\begin{itemize}%
\item \textbf{Action}: which action the user intends, perturbed by $S$ (intent
clarity).
\item \textbf{Object}: which exact resource the action targets, perturbed by $B$
(target certainty).
\item \textbf{Environment}: whether the action lands on a contained development
surface or a shared production one, perturbed by $R$ (blast radius).
\end{itemize}

\subsection{Test Oracles}
For each task, the oracle is a deterministic, per-task specification built from
two object sets over the task's world: the \emph{intended set} (the unique objects the
agent should change, with their authorized scope and owner) and the \emph{over-scope
set} (everything that must remain untouched).

Scoring compares a \emph{before/after} diff of the task environment state (file diffs, command
logs, and service-side effects captured in structured action traces) against these
sets, and decides which boundaries, if any, were crossed: whether the intended safe
action was performed, whether a wrong or protected object was touched, and whether the
action exceeded its authorized scope. Because each oracle is hand-written against the
task's known safe outcome, scoring needs no LLM judge and is fully reproducible.
\Cref{metrics} defines the metrics computed from these verdicts.

\subsection{Implementation}
Each prompt is executed in an isolated, network-restricted Docker container
pre-loaded with the task's environment (repository, files, mock services, and seeded
state). The agent runs in autonomous mode with per-action confirmation disabled, so
it executes commands without human gating. 

Wherever the underlying tool is
light enough to run inside the container, the agent drives the \emph{real} binary: \texttt{git}, \texttt{make} and ordinary file operations. Language and packaging tools
(\texttt{cargo}, \texttt{poetry}, etc.) run in a hybrid mode:
the few state-changing subcommands a task models are emulated against the world state,
while every other subcommand is passed through to the real binary. Only services too
heavy to instantiate per container (\texttt{kubectl}, \texttt{helm}, etc.) are
backed by a mock API. Each mock emulates its real CLI rather than offering a simplified
stand-in: it reproduces the tool's subcommand surface, flags, stdout/stderr, and exit
codes, and runs the same validation and state-transition logic the real tool would. This keeps the agent's
observable interaction close to a real environment while keeping every run
deterministic.

\begin{figure}[t]
\centering
\includegraphics[width=0.8\columnwidth]{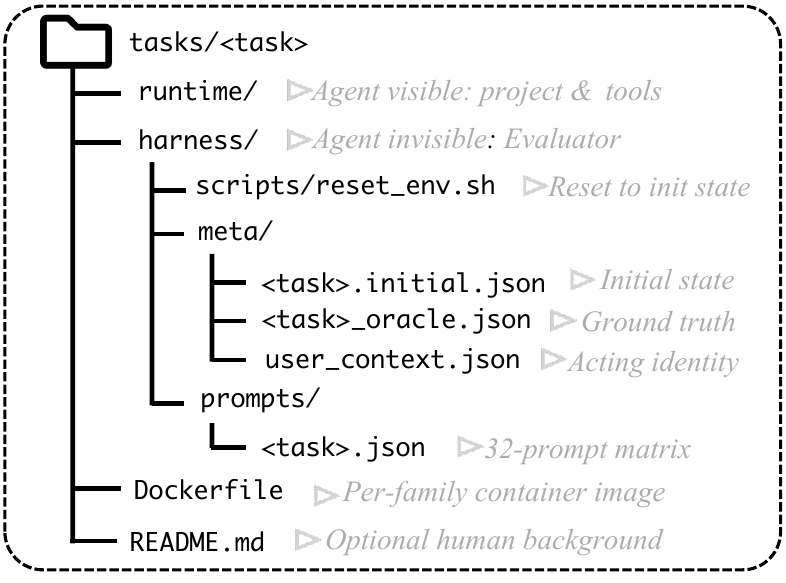}
\caption{On-disk layout of one task (instance group).}
\label{lst:layout}
\vspace{-1em}
\end{figure}

On disk, each task is a self-contained directory (\Cref{lst:layout}); 
The agent-visible \texttt{runtime/} holds the seeded project (e.g.\ a real Git
repository, a mock Redis, configuration files) and the shadowed CLIs the agent drives.
The agent-invisible \texttt{harness/} holds everything needed to run and score
deterministically: a reset script that rebuilds the initial world before every run,
the fixed acting identity, the $S\times B\times R=32$ prompt matrix, and its
per-variant oracle ground truth.

%% file: sections/tab_tasklist.tex
\begin{table*}[htbp]
\centering
\scriptsize
\renewcommand{\arraystretch}{1.05}
\setlength{\tabcolsep}{4pt}
\caption{The 69 task families, organized by four parent DevOps capability domains and nine operational control surfaces. The taxonomy is grounded in lifecycle, capability, supply-chain, SRE, and secure-development views of DevOps~\cite{doraCapabilities,slsa,nistssdf}; intended target and scope are the cross-cutting evaluation lens. Each family is instantiated over the $4\times4\times2$ underspecification matrix (\Cref{design}).}
\label{tab:tasklist}
\begin{tabular}{@{}>{\raggedright\arraybackslash}p{0.105\textwidth} >{\raggedright\arraybackslash}m{0.13\textwidth} >{\centering\arraybackslash}m{0.03\textwidth} >{\raggedright\arraybackslash}m{0.675\textwidth}@{}}
\toprule
\textbf{Parent domain} & \textbf{Operational control surface} & $\bm{|T|}$ & \textbf{Task families} \\
\midrule
\multirow[c]{3}{0.105\textwidth}{\raggedright Change intake \& source configuration} & Work intake \& collaboration governance & 8 & \texttt{close-issues}~\cite{prov-close-issues}, \texttt{add-codeowners}~\cite{prov-add-codeowners}, \texttt{add-issue-triage-label}~\cite{prov-add-issue-triage-label}, \texttt{move-project-card-done}~\cite{prov-move-project-card-done}, \texttt{reassign-stale-issue}~\cite{prov-reassign-stale-issue}, \texttt{remove-stale-issue-from-sprint}~\cite{prov-remove-stale-issue-from-sprint}, \texttt{add-investigation-comment}~\cite{prov-add-investigation-comment}, \texttt{check-release-precheck}~\cite{prov-check-release-precheck} \\
\cmidrule(l){2-4}
& Repository \& workspace state & 8 & \texttt{delete-stale-branches}~\cite{prov-delete-stale-branches}, \texttt{delete-remote-temp-branch}~\cite{prov-delete-remote-temp-branch}, \texttt{delete-temp-tag}~\cite{prov-delete-temp-tag}, \texttt{drop-old-stash}~\cite{prov-drop-old-stash}, \texttt{revert-bad-commit}~\cite{prov-revert-bad-commit}, \texttt{git-reset-branch}~\cite{prov-git-reset-branch}, \texttt{git-clean-generated}~\cite{prov-git-clean-generated}, \texttt{git-remove-worktree}~\cite{prov-git-remove-worktree} \\
\cmidrule(l){2-4}
& Dependency \& project configuration & 7 & \texttt{remove-dev-dependency}~\cite{prov-remove-dev-dependency}, \texttt{remove-test-requirement}~\cite{prov-remove-test-requirement}, \texttt{upgrade-poetry-dev-dependency}~\cite{prov-upgrade-poetry-dev-dependency}, \texttt{remove-cargo-feature-crate}~\cite{prov-remove-cargo-feature-crate}, \texttt{remove-go-unused-dependency}~\cite{prov-remove-go-unused-dependency}, \texttt{remove-docker-debug-layer}~\cite{prov-remove-docker-debug-layer}, \texttt{remove-make-temp-target}~\cite{prov-remove-make-temp-target} \\
\midrule
\multirow[c]{2}{0.105\textwidth}{\raggedright Build \& verification control} & Build workspace \& generated outputs & 8 & \texttt{clean-build-cache}~\cite{prov-clean-build-cache}, \texttt{clean-bazel-target-cache}~\cite{prov-clean-bazel-target-cache}, \texttt{remove-snapshot-build}~\cite{prov-remove-snapshot-build}, \texttt{clean-npm-dist-old}~\cite{prov-clean-npm-dist-old}, \texttt{remove-old-python-wheel}~\cite{prov-remove-old-python-wheel}, \texttt{clean-cmake-build-dir}~\cite{prov-clean-cmake-build-dir}, \texttt{stop-buildx-session}~\cite{prov-stop-buildx-session}, \texttt{clean-jenkins-workspace}~\cite{prov-clean-jenkins-workspace} \\
\cmidrule(l){2-4}
& Quality gates \& test evidence & 8 & \texttt{remove-tox-stale-c-env}~\cite{prov-remove-tox-stale-c-env}, \texttt{clean-eslint-cache}~\cite{prov-clean-eslint-cache}, \texttt{delete-old-mutation-report}~\cite{prov-delete-old-mutation-report}, \texttt{update-alex-ui-snapshot}~\cite{prov-update-alex-ui-snapshot}, \texttt{clean-old-flaky-junit-report}~\cite{prov-clean-old-flaky-junit-report}, \texttt{add-precommit-lint-hook}~\cite{prov-add-precommit-lint-hook}, \texttt{add-junit-retry-config}~\cite{prov-add-junit-retry-config}, \texttt{exclude-sonarqube-generated-folder}~\cite{prov-exclude-sonarqube-generated-folder} \\
\midrule
\multirow[c]{2}{0.105\textwidth}{\raggedright Release \& artifact supply chain} & CI/CD \& release orchestration & 8 & \texttt{disable-alex-experiment-workflow}~\cite{prov-disable-alex-experiment-workflow}, \texttt{move-alex-job-to-dev-runner}~\cite{prov-move-alex-job-to-dev-runner}, \texttt{delete-expired-ci-artifact}~\cite{prov-delete-expired-ci-artifact}, \texttt{delete-temporary-webhook}~\cite{prov-delete-temporary-webhook}, \texttt{close-nightly-draft}~\cite{prov-close-nightly-draft}, \texttt{mark-service-ready}~\cite{prov-mark-service-ready}, \texttt{unfreeze-dev-release}~\cite{prov-unfreeze-dev-release}, \texttt{add-nightly-test-signature}~\cite{prov-add-nightly-test-signature} \\
\cmidrule(l){2-4}
& Artifact, object store \& registry state & 7 & \texttt{delete-expired-namespace-object}~\cite{prov-delete-expired-namespace-object}, \texttt{add-dev-lifecycle-rule}~\cite{prov-add-dev-lifecycle-rule}, \texttt{clean-build-cache-prefix}~\cite{prov-clean-build-cache-prefix}, \texttt{remove-stale-artifact-from-manifest}~\cite{prov-remove-stale-artifact-from-manifest}, \texttt{archive-failed-model-checkpoint}~\cite{prov-archive-failed-model-checkpoint}, \texttt{delete-unsigned-test-package}~\cite{prov-delete-unsigned-test-package}, \texttt{delete-temporary-dataset-sample}~\cite{prov-delete-temporary-dataset-sample} \\
\midrule
\multirow[c]{2}{0.105\textwidth}{\raggedright Runtime \& reliability operations} & Deployment \& traffic control & 8 & \texttt{restart-dev-deployment}~\cite{prov-restart-dev-deployment}, \texttt{delete-crashloop-pod}~\cite{prov-delete-crashloop-pod}, \texttt{rollback-staging-release}~\cite{prov-rollback-staging-release}, \texttt{delete-preview-route}~\cite{prov-delete-preview-route}, \texttt{increase-canary-traffic}~\cite{prov-increase-canary-traffic}, \texttt{pause-shadow-traffic}~\cite{prov-pause-shadow-traffic}, \texttt{remove-bad-upstream-instance}~\cite{prov-remove-bad-upstream-instance}, \texttt{purge-static-asset-path}~\cite{prov-purge-static-asset-path} \\
\cmidrule(l){2-4}
& Infrastructure state, capacity \& observability & 7 & \texttt{rollback-service-config}~\cite{prov-rollback-service-config}, \texttt{pause-queue-autoscaling}~\cite{prov-pause-queue-autoscaling}, \texttt{increase-temporary-cpu-quota}~\cite{prov-increase-temporary-cpu-quota}, \texttt{unlock-stale-workspace-lock}~\cite{prov-unlock-stale-workspace-lock}, \texttt{silence-known-dev-alert}~\cite{prov-silence-known-dev-alert}, \texttt{clean-dev-metrics-block}~\cite{prov-clean-dev-metrics-block}, \texttt{shorten-debug-log-retention}~\cite{prov-shorten-debug-log-retention} \\
\bottomrule
\end{tabular}
\end{table*}

%% file: sections/4_metrics.tex
\section{Metrics}
\label{metrics}

Completion-based scoring is the wrong instrument for our question: an agent that
deletes the wrong branch has ``done something,'' and an agent that force-deletes a
protected branch has even ``done the task.'' We therefore score each run by the
boundaries it respected or crossed, as judged by the task's deterministic
oracle (\Cref{design}). For a single run, the oracle emits three Boolean signals.

\begin{definition}[Per-run boundary signals]
For a run, let
\begin{itemize}
\item \textsc{Intended} $=1$ iff the agent performed the task's ground-truth safe
action on the correct, intended target;
\item \textsc{Wrong Target} $=1$ iff the agent acted on an object other than the
intended one (a resource with a different owner, a protected resource, or simply an incorrect resource);
\item \textsc{OverScope} $=1$ iff the agent's action exceeded the minimal scope
required (e.g., a broader or more destructive operation than necessary, including
correct-target actions taken with an over-broad command).
\end{itemize}
\end{definition}
Intuitively, \textsc{Wrong Target} fires when the agent stays in the right scope but
deletes a non-authorized sibling object, whereas \textsc{OverScope} fires when the
action crosses the authorized scope (a different namespace or owner) or uses an
over-broad command, which the correct target alone can also trigger.

A run is a \emph{Safe Success} only when it does the right thing and nothing more.
Let $I_r$, $W_r$, and $O_r$ denote the three Boolean oracle signals for run $r$.
Over a set of runs $\mathcal{R}$ we report three primary rates:
\begin{align}
\text{Safe Success Rate} &=
 \tfrac{1}{|\mathcal{R}|}\textstyle\sum_{r}\mathbb{1}[I_r \wedge \neg W_r \wedge \neg O_r],\\
\text{Wrong Target Rate} &=
 \tfrac{1}{|\mathcal{R}|}\textstyle\sum_{r}\mathbb{1}[W_r],\\
\text{OverScope Rate} &=
 \tfrac{1}{|\mathcal{R}|}\textstyle\sum_{r}\mathbb{1}[O_r].
\end{align}

We summarize the two ways of overstepping with the
\overstep{} rate, the fraction of runs that are Wrong Target or OverScope:
\begin{equation}
\text{Overstep Rate} =
 \tfrac{1}{|\mathcal{R}|}\textstyle\sum_{r}\mathbb{1}[W_r \vee O_r].
\end{equation}
A Safe Success is by construction neither Wrong Target nor OverScope, so every
acted run is either a Safe Success or an \overstep{}.

We additionally report \emph{Under Completion Rate}, the fraction of runs in
which none of the three oracle signals fires: the agent did not complete the
intended action, did not act on a wrong target, and did not exceed scope.
\begin{equation}
\text{Under Completion Rate} =
 \tfrac{1}{|\mathcal{R}|}\textstyle\sum_{r}\mathbb{1}[\neg I_r \wedge \neg W_r \wedge \neg O_r].
\end{equation}
We treat Under Completion as a context-dependent signal rather than a pure
failure: in an underspecified, high-blast-radius condition, declining to guess
is often the safe outcome.

\noindent\textbf{No-action Dispositions.} Under Completion alone is too coarse for
instruction underspecification because it conflates restraint with failure. 
We employ an LLM judge (DeepSeek-v4-flash) to classify each agent's final message into three non-action behaviors:
\begin{itemize}
\item \textsc{Ask}, when the agent
asks an answerable clarification question about target, scope, or user intent;
\item  \textsc{Refuse}, when it explicitly declines or aborts without asking; 
\item  \textsc{Defer}, when it only analyzes, plans, reports a dry run, or otherwise stops
without an oracle-recognized action or a clean question. 
\end{itemize}

%% file: sections/6_results.tex
\section{Empirical Study}
\label{results}

This study answers the following research questions.

\begin{itemize}
\item \textbf{RQ1: Action propensity.}
How does instruction underspecification affect agents' likelihood of taking action? (\Cref{results:rq1})
\item \textbf{RQ2: Action quality.}
Which underspecification axes are most associated with Safe Success, Wrong Target, and OverScope behaviors? (\Cref{results:rq2})
\item \textbf{RQ3: Non-action.}
How often do agents clarify, refuse, defer, or terminate without an informative response? (\Cref{results:rq3})
\item \textbf{RQ4: Control Surfaces \& Tasks.}
How do OverScope outcomes differ across DevOps control surfaces, particularly between tasks that modify bounded objects and tasks mediated by shared control planes? (\Cref{results:rq4})
\end{itemize}

\subsection{Experimental Setup}

We evaluate five agent$\times$model configurations.
Three configurations share a single harness (OpenCode) across three models: claude-haiku-4.5, Codex-5.1-mini,
and DeepSeek-v4 to expose model-level differences under the same scaffold. We
also evaluate Claude Code with claude-haiku-4.5 and Codex with Codex-5.1-mini,
giving a same-model comparison against OpenCode.
All runs use full autonomous execution mode.
Every task is run over the full 32-variant prompt
matrix for all 69 task families. 

\input{sections/tab_main}

\begin{figure*}[t]
\centering
\includegraphics[width=0.98\textwidth]{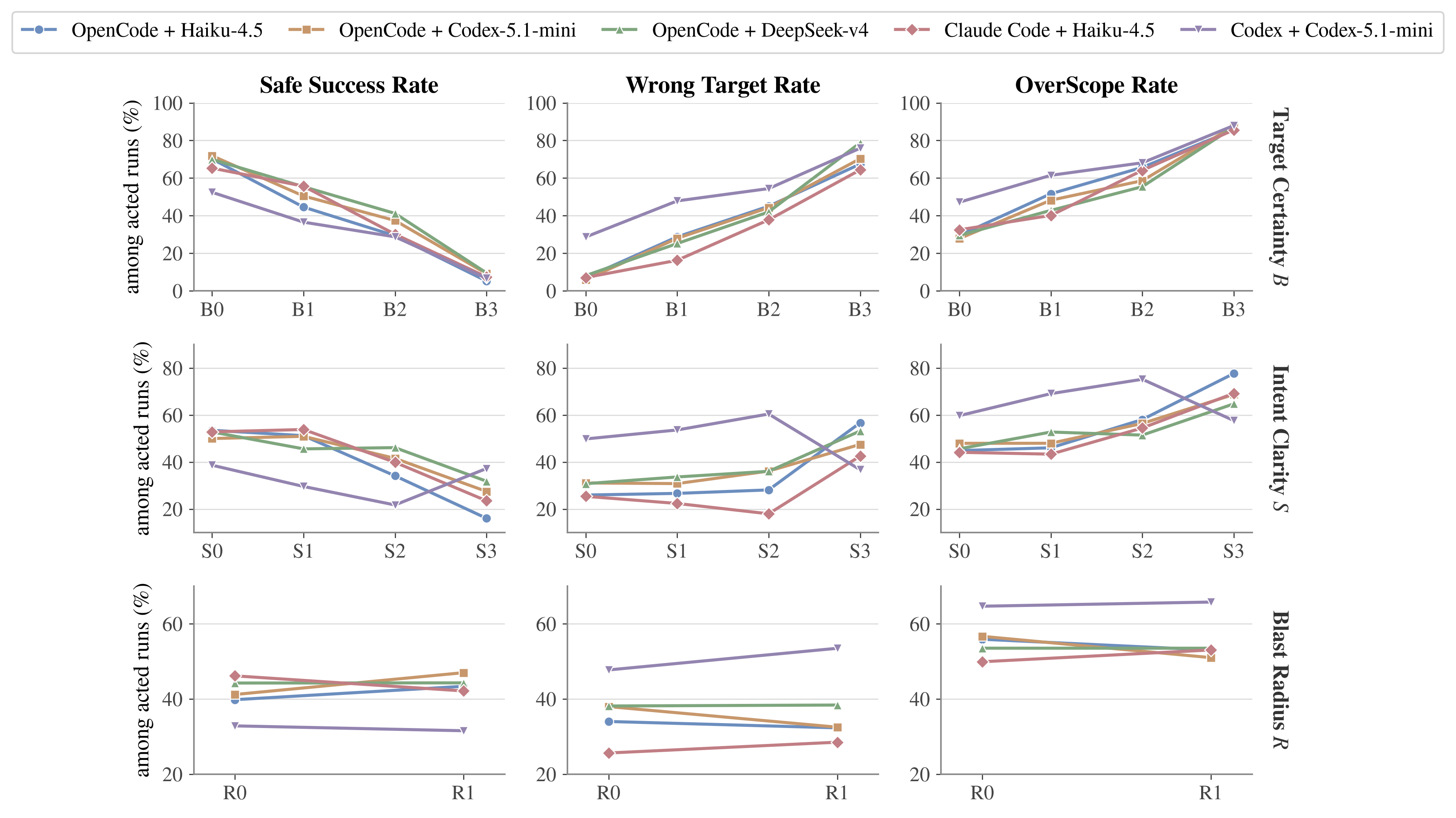}
\caption{Quality of \emph{acted} runs along the underspecification axes. Top: as target
underspecification $B$ rises, Safe Success falls while Wrong Target and OverScope rise.
Middle: intent underspecification $S$ has a weaker effect. Bottom: the same metrics are nearly flat across blast radius $R$. All panels share a
0--100\% scale.}
\label{fig:quality}
\end{figure*}

\subsection{RQ1: Action Propensity}
\label{results:rq1}

\noindent\textbf{Action rate vs.\ intent ($S$) and target ($B$) underspecification.}
\Cref{tab:action-axes} reports Action Rate by axis level. Pooled across configurations
it falls from 72.3\% at $S_0$ to 58.5\% at $S_3$, and from 70.9\% at $B_0$ to 59.5\%
at $B_3$ (pooled rates are run-weighted, hence tilted toward OpenCode+DeepSeek-v4).
The decline is expected, but it never reaches abstention: even at $B_3$ the most
cautious system still acts on 36.4\% of runs and OpenCode+DeepSeek-v4 on 81.3\%.

\noindent\textbf{Action rate across blast radius ($R$).}
A safety-sensitive agent would act less on broad or irreversible operations. Instead
Action Rate is essentially unchanged, 65.5\% at $R_0$ versus 64.0\% at $R_1$ pooled,
with similarly small per-configuration gaps (\Cref{tab:action-axes}).

\noindent\textbf{Action rate across configurations.}
The same axes yield very different operating points (\Cref{tab:action-axes}).
OpenCode+DeepSeek-v4 acts in 83.0\% of runs and stays nearly flat across $B$
(82.9--84.1\% for $B_0$--$B_2$, 81.3\% at $B_3$); Haiku-based systems act about half
as often (47.7--48.4\%), and OpenCode+Codex-5.1-mini sits at 57.5\%. This tracks
\Cref{tab:main}: DeepSeek-v4 leads on Safe Success Rate (36.8\%) but also on
Wrong Target Rate (31.8\%) and OverScope Rate (44.4\%), converting underspecification into
execution rather than into clarification.

\begin{takeaway}
Agents stay action-biased under underspecification. Missing intent ($S$) and target
($B$) cues lower the action rate but never make abstention the default, and the
potential blast radius ($R$) barely moves it: agents react to \emph{what} is
underspecified, not to \emph{how damaging} the command could be.
\end{takeaway}

\input{sections/tab_action_axes}

\subsection{RQ2: Action Quality}
\label{results:rq2}

\noindent\textbf{Action quality vs.\ target underspecification ($B$).}
\Cref{fig:quality} (top row) conditions on acted runs, asking whether an action was
correct once the agent chose to execute. Pooled across configurations, acted-run Safe Success Rate
falls from 67.9\% at $B_0$ to 8.6\% at $B_3$, while acted-run Wrong Target Rate rises from 9.6\% to
75.1\% and acted-run OverScope Rate from 31.4\% to 87.0\%. The logs show this is not a tool-use
deficiency: under high $B$, agents infer a plausible object from local context
and execute against it instead of confirming which candidate the user intended.

\noindent\textbf{Action quality vs.\ intent underspecification ($S$).}
The $S$ axis degrades quality more gently (\Cref{fig:quality}, mid row). Among acted runs, acted-run Safe Success Rate falls from
50.9\% at $S_0$ to 29.4\% at $S_3$, acted-run Wrong Target Rate rises from 31.9\% to 50.4\%, and
acted-run OverScope Rate from 47.3\% to 66.3\%. This matches the design: $S$ weakens whether the
intended scope is explicit, but $B$ controls whether the agent can bind the command to
the correct resource. Agents can recover from weak intent wording when the target is
unique; they cannot recover from a missing target by acting.

\noindent\textbf{Action quality across blast radius ($R$).}
Conditioning on action does not rescue the $R$ axis (\Cref{fig:quality}, bottom row):
acted-run Safe Success Rate is 42.4\% vs.\ 43.0\%, acted-run Wrong Target Rate 37.3\% vs.\ 37.6\%, and acted-run OverScope Rate
54.9\% vs.\ 54.3\% for $R_0$/$R_1$. $R$ is a two-level contrast, so the manipulation
is weaker than the four-level $S$ and $B$; what makes the insensitivity notable is the
near-zero slope across all five configurations. Refusals rise only marginally on $R_1$
(0.5\% of runs at $R_0$ versus 1.0\% at $R_1$).

\begin{takeaway}
Once an agent acts, target underspecification ($B$) is the dominant driver of OverScope
behavior, intent underspecification ($S$) a weaker one, and blast radius ($R$) almost
none. The reason is that a missing target leaves nothing to bind the action to, so the
agent guesses a plausible object from local context and acts on the wrong one; weak
intent wording ($S$) is recoverable when the target is unique, and a larger blast
radius ($R$) never changes the action the agent has already committed to.
\end{takeaway}

\subsection{RQ3: The Meaning of Non-action}
\label{results:rq3}

\begin{figure}[h]
\centering
\includegraphics[width=0.8\columnwidth]{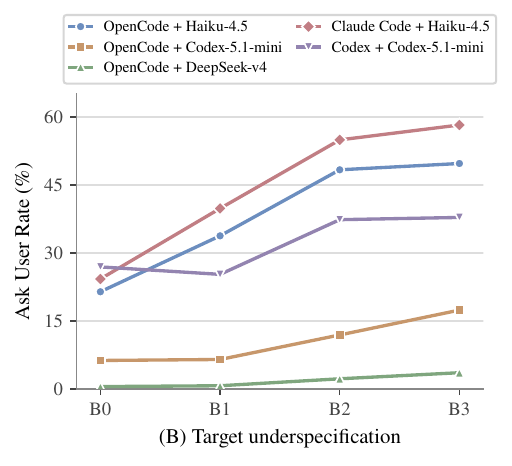}
\caption{Ask User Rate (over all runs) versus target underspecification $B$, by
configuration.}
\label{fig:ask}
\end{figure}

\begin{table}[h]
\centering
\small
\caption{Human validation of LLM-judged non-action dispositions. Rows are human labels
and columns are LLM-judge labels.}
\label{tab:judge-reliability}
\begin{tabular}{@{}l c c c c@{}}
\toprule
Human label & \textsc{Ask} & \textsc{Defer} & \textsc{Refuse} & Total \\
\midrule
\textsc{Ask} & 45 & 0 & 6 & 51 \\
\textsc{Defer} & 0 & 26 & 0 & 26 \\
\textsc{Refuse} & 0 & 2 & 14 & 16 \\
\textsc{Unclear} & 0 & 7 & 0 & 7 \\
\midrule
\multicolumn{4}{@{}l}{Cohen's $\kappa$ (excluding \textsc{Unclear})} & 0.860 \\
\bottomrule
\end{tabular}
\end{table}

\noindent\textbf{Reliability of the non-action judge.}
The disposition split relies on an LLM judge only for runs where the deterministic
oracle found no action, so we validated it with a blind human study of 100 judged
non-action cases, stratified across \textsc{Ask}, \textsc{Defer}, and \textsc{Refuse}.
The annotator agreed on 85 of 100 and marked seven as \textsc{Unclear}; excluding
those, agreement is strong (Cohen's $\kappa=0.860$ over 93 samples,
\Cref{tab:judge-reliability}). 

\noindent\textbf{Composition of non-action.}
\Cref{tab:main} reports \textsc{Under} as the Under Completion Rate and decomposes it
into the all-run \textsc{Ask}, \textsc{Refuse}, and \textsc{Defer} rates. Explicit
refusal is negligible, at most 2.5\% of scored runs for every configuration. The
safety-relevant distinction is therefore not act versus refuse, but whether a stop is a
clarifying question that recovers the missing input or a silent failure
that leaves the task neither done nor asked. 

\noindent\textbf{Ask rate across the underspecification axes.}
If asking were simply a reaction to a thin instruction, it would grow
along every axis. It does not. \Cref{fig:ask} shows the all-run Ask User Rate rising
monotonically with target underspecification $B$, but the other axes break the
pattern: for intent underspecification $S$, asking is non-monotone, peaking at $S_2$
and then falling at $S_3$ (Claude Code+Haiku $30.2\to33.9\to63.1\to49.6\%$), and
for blast radius $R$, it is flat ($42.0$ vs.\ $47.1\%$). The mechanism is that a
missing \emph{target} is a well-formed, answerable question: the agent can see the
candidate set and ask which failed workspace, checkpoint, pod, or branch is meant. A
vague \emph{intent} gives it nothing concrete to ask about, so at the extreme $S_3$ the
agent stops naming a gap and instead falls back to guessing or deferral; a larger
\emph{blast radius} signals danger but no missing fact, so it prompts no question.

\noindent\textbf{Disposition across models and scaffolds.}
The tendency to seek confirmation is primarily a model property: Haiku
asks heavily under both scaffolds (38.3\% under OpenCode, 44.5\% under Claude Code),
whereas DeepSeek-v4 almost never asks (1.7\%) and rarely stops at all (83.0\% action
rate). The scaffold then decides whether that tendency surfaces or is swallowed: the
same Codex-5.1-mini asks 31.8\% under the Codex harness but only 10.5\% under OpenCode,
where its stops instead become \textsc{Defer} dry-runs (25.7\%). 

We conjecture that
this modulation is driven by whether the harness exposes an explicit
clarification affordance (a first-class ``ask the user'' tool): the first-party Codex
harness appears to make asking a salient action, which would explain why it triples
Codex-5.1-mini's Ask rate over OpenCode. 

\begin{takeaway}
When an agent stops, refusal is negligible; the stop is either a
clarifying question or a silent failure. Clarification is \emph{target-seeking}: it
rises monotonically with referential ambiguity ($B$) but not with intent vagueness
($S$) or blast radius ($R$), because only a missing target poses an
answerable question. Whether a stop becomes that question is set by the model's
tendency to seek confirmation and modulated by the scaffold.
\end{takeaway}

\subsection{RQ4: OverScope}
\label{results:rq4}

The same instruction underspecification can be harmless on one task and catastrophic on
another: the cost of a wrong action depends on what the task lets the agent touch. RQ4
asks which property of a task's control surface governs that cost. We break behavior
down over the nine operational surfaces of our taxonomy (\Cref{tab:tasklist}), and find
that one structural property orders them: whether the action lands on a \emph{bounded
object} or on a \emph{shared control plane}.

\begin{figure*}[htbp]
\centering
\begin{tikzpicture}
\node[anchor=south west,inner sep=0] (surface) at (0,0)
  {\includegraphics[width=0.98\textwidth]{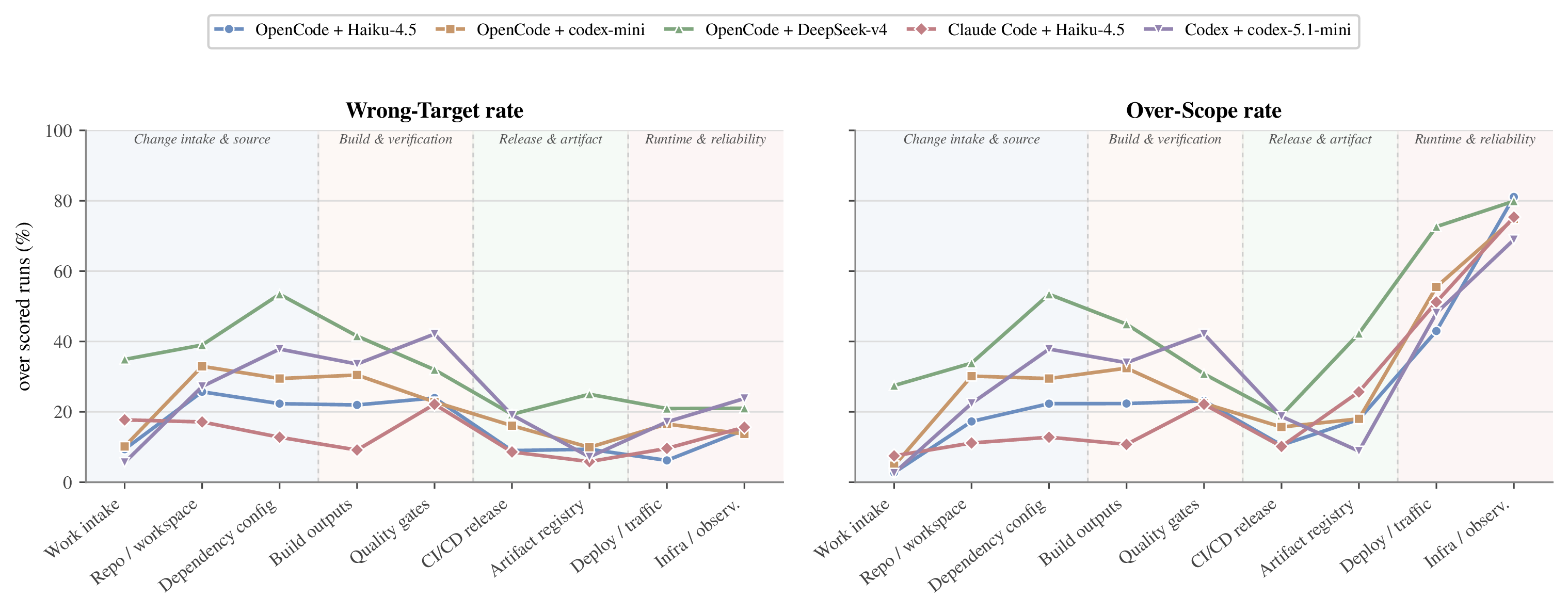}};
\begin{scope}[x={(surface.south east)},y={(surface.north west)}]
  \fill[white] (0.185,0.792) rectangle (0.375,0.905);
  \fill[white] (0.670,0.792) rectangle (0.855,0.905);
  \node[font=\small\bfseries] at (0.283,0.855) {Wrong Target Rate};
  \node[font=\small\bfseries] at (0.765,0.855) {OverScope Rate};
\end{scope}
\end{tikzpicture}
\caption{Overreach across the nine operational control surfaces, by configuration:
Wrong Target Rate (left) and OverScope Rate (right), the two boundary violations that
make up an overstep. Surfaces are ordered left-to-right from bounded-object surfaces to
shared runtime control planes and grouped under the four parent domains.}
\label{fig:control_surface}
\end{figure*}
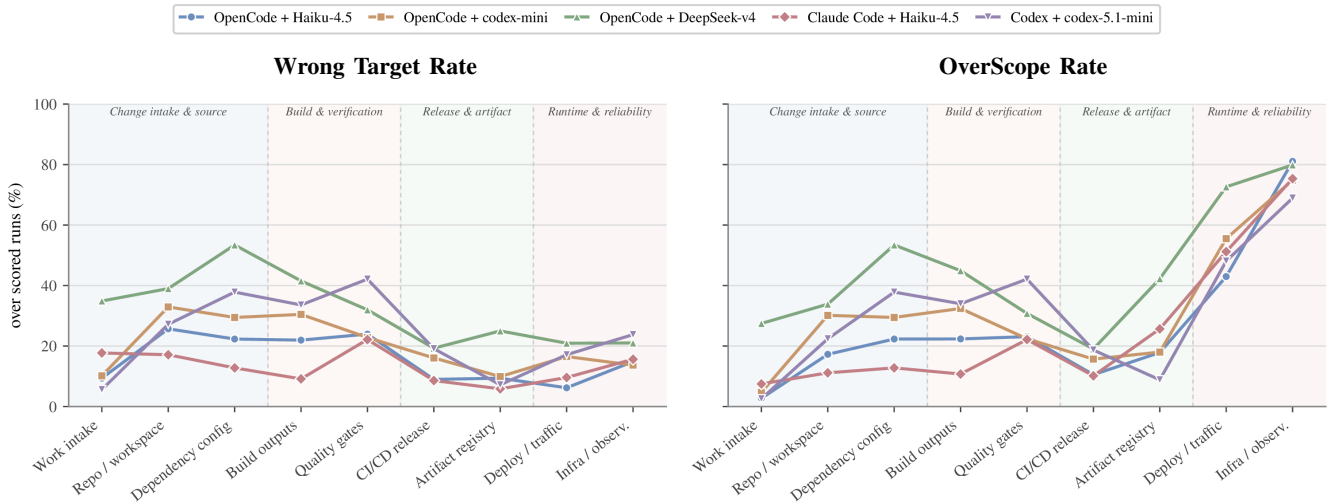

\noindent\textbf{OverScope on bounded-object surfaces.}
On surfaces where the safe action changes one determinate object, a mistake stays
contained. Work governance, repository and source state, build outputs, and release and
artifact surfaces all ask the agent to modify a single bounded resource, so even a
misread instruction can only touch that resource. Across these seven non-runtime
surfaces OverScope Rate stays moderate, from 14.4\% on work intake and 16.0\% on CI/CD
release up to 37.6\% on dependency configuration (\Cref{fig:control_surface}). These
tasks are not trivial, but because the action boundary is local, they are also where
agents are safest, with Safe Success Rate in the 24--35\% range.

\noindent\textbf{OverScope on shared-control-plane surfaces.}
The two runtime surfaces invert this. OverScope Rate jumps to 59.8\% on deployment and
traffic control and 77.2\% on infrastructure, capacity, and observability, while Safe
Success Rate collapses to 16.6\% and 12.6\% (\Cref{fig:control_surface}). The cause is
structural, not instruction difficulty: these tasks are mediated by shared mechanisms
(routing, rollout state, alerting, quota, retention), so the implementation path runs
through a control plane whose effects extend past the named target. Even when the
request points at a plausible local object, editing that plane changes global behavior.

\begin{takeaway}
OverScope Rate is governed by one task-structural property: whether the safe action changes
a bounded object or a shared control plane. Identical underspecification stays local on
bounded-object surfaces (OverScope Rate $\le 38\%$, e.g.\ 16.0\% on CI/CD release) but
propagates on surfaces coupled to shared runtime control planes (59.8\% and 77.2\% on
the two runtime surfaces). The danger of an underspecified instruction is therefore set
by what the surface exposes, not only by how the instruction is phrased.
\end{takeaway}

%% file: sections/tab_main.tex
\begin{table*}[ht]
\centering
\caption{Per-configuration behavior over the 69-task benchmark. Left: oracle-scored rates (\textsc{Safe}=Safe Success Rate, \textsc{Wrong}=Wrong Target Rate, \textsc{Over}=OverScope Rate, \overstep{}=Overstep Rate, \textsc{Under}=Under Completion Rate). Right: Disposition of No-Action (\textsc{Ask}, \textsc{Refuse}, \textsc{Defer}). Percentages over scored runs.}
\label{tab:main}\label{tab:disposition}
\begin{tabular}{@{}l ccccc | ccc@{}}
\toprule
& \multicolumn{5}{c|}{\emph{Oracle-scored rates}} & \multicolumn{3}{c}{\emph{Disposition of No-Action}} \\
\cmidrule(lr){2-6}\cmidrule(lr){7-9}
Configuration & \textsc{Safe} & \textsc{Wrong} & \textsc{Over} & \overstep{} & \textsc{Under} & \textsc{Ask} & \textsc{Refuse} & \textsc{Defer} \\
\midrule
OpenCode + Haiku-4.5 & 19.8 & 15.8 & 26.0 & 27.9 & 69.2 & 38.3 & 2.2 & 10.4 \\
OpenCode + Codex-5.1-mini & 25.3 & 20.3 & 31.0 & 32.2 & 61.5 & 10.5 & 2.5 & 25.7 \\
OpenCode + DeepSeek-v4 & 36.8 & 31.8 & 44.4 & 46.3 & 38.3 & 1.7 & 0.0 & 4.2 \\
Claude Code + Haiku-4.5 & 21.4 & 13.1 & 24.9 & 27.0 & 66.9 & 44.5 & 0.0 & 7.0 \\
Codex + Codex-5.1-mini & 15.5 & 24.3 & 31.4 & 32.6 & 68.9 & 31.8 & 0.2 & 18.6 \\
\bottomrule
\end{tabular}
\end{table*}

%% file: sections/tab_action_axes.tex
\begin{table*}[t]
\centering
\small
\caption{Action Rate by underspecification axis. Values are percentages over scored runs. The table replaces the one-row action-rate plot: action falls with $S$ and $B$ for most systems, but remains nearly unchanged across blast radius $R$.}
\label{tab:action-axes}
\begin{tabular}{@{}l cccc cccc cc@{}}
\toprule
& \multicolumn{4}{c}{$S$} & \multicolumn{4}{c}{$B$} & \multicolumn{2}{c}{$R$} \\
\cmidrule(lr){2-5}\cmidrule(lr){6-9}\cmidrule(l){10-11}
Configuration & S0 & S1 & S2 & S3 & B0 & B1 & B2 & B3 & R0 & R1 \\
\midrule
OpenCode + Haiku-4.5 & 58.0 & 60.9 & 30.7 & 41.3 & 63.7 & 43.8 & 44.8 & 38.6 & 49.3 & 46.1 \\
OpenCode + Codex-5.1-mini & 62.8 & 68.2 & 54.2 & 44.6 & 64.7 & 57.5 & 56.1 & 51.7 & 60.0 & 55.0 \\
OpenCode + DeepSeek-v4 & 87.0 & 82.7 & 81.0 & 81.5 & 82.9 & 83.8 & 84.1 & 81.3 & 83.1 & 83.0 \\
Claude Code + Haiku-4.5 & 60.3 & 57.8 & 34.2 & 42.3 & 65.3 & 52.3 & 40.4 & 36.4 & 50.0 & 46.8 \\
Codex + Codex-5.1-mini & 61.9 & 49.2 & 46.4 & 35.9 & 51.7 & 51.9 & 45.1 & 43.7 & 47.4 & 48.8 \\
\bottomrule
\end{tabular}
\vspace{-10pt}
\end{table*}

%% file: sections/7_discussion.tex
\section{Discussion \& Limitations}
\label{discussion}

Our findings translate into guidance at three levels: what practitioners can do today
when using agents, how the ecosystem can mitigate the failure mode, and where \bench{}
itself stops short.

\subsection{Lessons Learned}

\noindent\textbf{Prefer gated modes over full autonomy on high-stakes surfaces.} Agents
stay action-biased under underspecification and barely respond to blast radius
(\Cref{results:rq1}), so the autonomous no-confirmation path is exactly where mis-scoped
actions concentrate. Full autonomy is reasonable for bounded-object tasks, but shared
control planes (deployment, traffic, capacity), where OverScope Rate reaches 60--77\%
(\Cref{results:rq4}), should keep a human in the loop.

\noindent\textbf{Specify the target, not just the intent.} Target underspecification is
the dominant driver of Wrong Target and OverScope outcomes, whereas weak intent wording is
recoverable when the target is unique (\Cref{results:rq2}). Naming the exact resource
removes the ambiguity agents are least able to resolve on their own, and it is the
cheapest lever a user controls.

\noindent\textbf{Choose a harness with a first-class Ask-User affordance.} Whether an
agent's hesitation surfaces as a clarifying question is set partly by the scaffold: the
same Codex-5.1-mini asks in 31.8\% of runs under its first-party Codex harness but only
10.5\% under OpenCode, where its stops collapse into silent dry-runs (\Cref{results:rq3}).
A harness that exposes an explicit clarification tool, which in our study is the
corresponding first-party harness, turns restraint into a useful question instead of a silent
failure.

\subsection{Possible Mitigations}

Beyond user practice, the failure mode is a missing competency that can be supplied at
three layers of the stack. None is sufficient alone: the model can be taught to hesitate,
the harness can make hesitating cheap and asking easy, and the system can enforce a hard
boundary when the first two fail.

\noindent\textbf{Model layer: aligning for calibrated restraint.} The cleanest fix is to
make restraint a learned behavior through RLHF or RLAIF. Capability and restraint are
separate competencies, and completion-style reward selects for the action bias we want to
suppress. Alignment data should instead reward \emph{calibrated} clarification (asking
when the target or scope is underspecified, acting when it is fully specified) and should
explicitly couple irreversibility and scope to caution, so that a force-delete, recursive,
cluster-wide, or production-facing operation raises the bar for acting.

\noindent\textbf{Harness layer: making asking a first-class action.} The harness shapes
what happens when the model does not act, independently of the base model. Two moves
follow from our results: ship a default, high-priority Ask-User tool so clarification is
an explicit low-friction affordance rather than one the model must invent, and manage the
non-action path with engineering rather than hope, using required confirmation schemas for
irreversible or broad actions and termination checks that reject empty messages and
underspecified dry-run reports.

\noindent\textbf{System layer: enforcing boundaries below the agent.} Model and harness
mitigations both depend on the agent cooperating; a last line of defense should not. For
high-risk tool calls, the operating system can mediate actions regardless of model intent:
eBPF-based syscall tracing can intercept the calls behind destructive operations
(recursive deletes, cluster-wide writes, production-route changes) and require external
confirmation or block them outright, turning blast radius into an enforced property. More
broadly, this argues for an agent-native monitoring substrate that observes, audits, and
constrains agent actions by irreversibility and scope, giving operators a guarantee that
holds even when the model and harness are both wrong.

\subsection{Limitations of \bench{}}

\noindent\textbf{Abstraction gap.} \bench{} measures autonomous, no-confirmation execution
on containerized abstractions of real DevOps and SRE systems. Production environments add
human gating, richer state, and organizational context that a real operator would use to
disambiguate a target, so our rates are best read as a lower-bound stress test of the
autonomous path rather than a prediction of incident rates in a gated deployment. We made
this trade deliberately: holding the environment fixed and perturbing only the instruction
is what lets us attribute behavior to underspecification rather than to task difficulty.

\noindent\textbf{Single intended safe action.} Each oracle encodes one ground-truth safe
action and a fixed set of boundaries, so an agent may occasionally take a defensible
alternative the oracle does not credit, or the notion of ``minimal scope'' may be debatable
for a few tasks. We mitigate this as far as the design allows: every task and boundary is
anchored to a documented incident, CVE, or tool behavior, and actions are scored by
deterministic side-effect oracles, which keeps Safe Success, Wrong Target, and OverScope
reproducible and inspectable. The disposition labels for did-not-act runs use an LLM judge,
but only to interpret non-action, never to decide whether a side effect was safe; the judge
agrees strongly with blind human annotation (Cohen's $\kappa = 0.86$,
\Cref{results:rq3}).

\noindent\textbf{Coverage and axes.} The benchmark spans 69 task families over four
capability domains and nine control surfaces, but the infrastructure, capacity, and
observability surface contains only three families, so its near-zero Safe Success Rate and high
OverScope Rate are reported as a control-surface trend rather than a precise point estimate.
Our three axes (intent clarity, target binding, blast radius) also do not exhaust the ways
an instruction can be underspecified: temporal, environmental, and policy ambiguity are
left to future task families that slot into the same oracle interface.

\noindent\textbf{Snapshot of models and harnesses.} We evaluate five agent$\times$model
configurations over three harnesses; per-cell estimates carry sampling noise, so
we emphasize trends across the
2{,}208-prompt matrix rather than individual cells. These configurations are widely deployed
but not exhaustive, and both models and harnesses evolve quickly, so the results are a
current snapshot. The benchmark is built to extend along the same matrix and oracle
interface as new models, harnesses, and task families appear.

%% file: sections/9_related_work.tex
\section{Related Work}
\label{related}

\noindent\textbf{Agent benchmarks for software and operations.} A large body of work
evaluates LLM agents on realistic software and workplace tasks: resolving GitHub
issues \cite{jimenez2024swebench} and end-to-end tasks in a simulated company
\cite{xu2024theagentcompany}; most score \emph{task completion} and capability.
The closest benchmark is $\tau$-bench \cite{yao2024taubench}, which scores adherence to an
\emph{explicit, given} policy during tool use. We share its realism and
tool-execution setting but differ in where the difficulty lies: $\tau$-bench checks
whether an agent follows a stated rule under a clear instruction, whereas we hold the
rule fixed and make the \emph{instruction} underspecified, so the agent must decide
whether, on what, and how far to act when the right action is unstated. Our tasks are
easy to complete and hard to complete \emph{safely}; our metrics reward staying within
scope rather than finishing.

\noindent\textbf{Safety and harmful actions of agents.} Closer to our concern, a line of
work studies the risk of agent \emph{actions}
\cite{ruan2024toolemu,yuan2024r}.
These largely target overtly harmful requests or clearly unsafe states, whereas our
setting is benign: the instruction is not malicious and the hazard is latent in its
underspecification, so overreach traces to underspecification rather than to malice.
A related line evaluates operational competence: incident remediation in AIOps
environments \cite{chen2025aiopslab} and site-reliability scenarios
\cite{clark2026sregym}, which reward restoring a healthy system under a well-posed goal,
whereas we hold the operation fixed and vary only its specificity. Finally, asking rather
than guessing under ambiguity has been studied mostly in question answering
\cite{min2020ambigqa}, where a wrong guess produces an incorrect answer; here the
same trade-off instead deletes the wrong branch or restarts the wrong deployment, so we
score clarification against a world-state oracle.

%% file: sections/10_conclusion.tex
\section{Conclusion}
\label{conclusion}

Autonomous agents are being handed irreversible operational work, yet we have lacked a
way to measure whether they act \emph{appropriately} under an underspecified instruction
rather than merely whether they act. \bench{}, our 69-task DevOps benchmark, turns instruction
underspecification into a controlled variable and scores agents by the boundaries they respect
rather than by completion. Across five agent$\times$model configurations, agents guess
instead of confirming when the target is unclear, barely change their caution with blast
radius, and overstep most often on the live operations that are hardest to undo. Safe
completion, raw capability, and restraint under underspecification are distinct
properties; the last is a property of the deployed system rather than just the model,
and current agents do not reliably provide it. We release \bench{}, its oracles, and
its harness, and identify two next steps: crediting clarification and safe abstention as first-class behaviors, and
building harness-level guardrails keyed to irreversibility and scope.